\documentclass[twocolumn,preprintnumbers,amsmath,amssymb]{revtex4}
\usepackage{graphicx}
\usepackage{dcolumn}
\usepackage{epsfig}
\usepackage{bm}
\def\be{\begin{equation}}
\def\ee{\end{equation}}
\def\bea{\begin{eqnarray}}
\def\eea{\end{eqnarray}}
\def\nnb{\nonumber}

\begin{document}
\preprint{ CERN-PH-TH/2011-089}

\title{Transition Radiation by Neutrinos}

\author{A. N. Ioannisian$^{a,b,c}$, D. A. Ioannisian$^{c,d}$, N. A.  Kazarian${^c}$, }
 \affiliation{
$^a$ Yerevan Physics Institute, Alikhanian Br.\ 2, 375036 Yerevan,
Armenia\\
$^b$ CERN, Theory Division, CH-1211 Geneva 23, Switzerland \\
$^c$ Institute for Theoretical Physics and Modeling, 375036, Yerevan, Armenia \\
$^d$ Physics Department, Yerevan State Univesrity, 1 Alex Manoogian, Armenia }

\begin{abstract}
We calculate the transition radiation process $\nu \to \nu \gamma$
at an interface of two media. The neutrinos are taken to be with only
standard-model couplings. The medium fulfills the dual purpose of
inducing an effective neutrino-photon vertex and of modifying the
photon dispersion relation. The transition radiation occurs when at 
least one of those quantities  have different values in different media.
The neutrino mass is ignored due to its negligible contribution.  
We  present a result for the probability of the transition radiation 
which is both accurate and analytic.
For $E_\nu =1$MeV neutrino crossing  polyethylene-vacuum interface the 
transition radiation probability is about $10^{-39}$ and the energy 
intensity is about $10^{-34}$eV. At the surface of the neutron stars the transition radiation probability may be $\sim 10^{-20}$. Our result on three orders of magnitude is larger than the results of previous calculations.\\
\end{abstract}


\maketitle

{\it Introduction.-}
In many astrophysical environments the absorption, emission, or
scattering of neutrinos occurs in media, in the presence of 
magnetic fields \cite{Raffelt:1996wa} or at the interface 
of two media. Of particular conceptual interest
are those reactions which have no counterpart in vacuum, notably
the plasmon decay $\gamma\to\bar\nu\nu$, the Cherenkov and
transition radiation processes $\nu\to\nu\gamma$. These reactions
do not occur in vacuum because they are kinematically forbidden
and because neutrinos do not couple to photons. In the presence of
a media (or a magnetic field), neutrinos acquire an effective
coupling to photons by virtue of intermediate charged particles.
In uniform media (or external field) the dispersion relations are
modified of all particles so that phase space is opened for
neutrino-photon reactions of the type $1\to 2+3$. The violation of
the translational invariance at the direction from one media into
another leads to the non conservation of thr  momentum at the same
direction so that transition radiation becomes kinematically
allowed.

The theory of the transition radiation by charged particle  has been diveloped in \cite{Ginzburg:1945zz}\cite{Garibyan:1959}.
It those articles authors used classical theory of electrodynamics. In \cite{Garibyan:1960} 
the quantum field theory was used for describing the phenomenon. 
The neutrinos have very tiny masses. 
Therefore one has to use the quantum field theory approach in order to study transition radiation by neutrinos.

The plasma process $\gamma\to\bar\nu\nu$ was first studied in \cite{Adams:1963zz}. The
\hbox{$\nu$-$\gamma$}-coupling is enabled by the presence of the
electrons of the background medium, and the process is
kinematically allowed because the photons acquire essentially an
effective mass. The plasma process is the dominant source for
neutrinos in many types of stars and thus is of great practical
importance in astrophysics~\cite{Raffelt:1996wa}. 
In a plasma there are electromagnetic excitations, namely, the longitudinal plasmons, $\gamma_L $, which four-momentum are space like for certain energies. The Cherenkov decay $\nu \to \nu \gamma_L$ was studied in \cite{Oraevsky:1987cu}.

The presence of a magnetic field induces an effective
$\nu$-$\gamma$-coupling. The Cherenkov decay in a magnetic field
was calculated in  \cite{Ioannisian:1996pn}. The $\gamma\to\bar\nu
\nu$ decay rate was calculated in \cite{DeRaad}, assuming that phase space is
opened by a suitable medium- or field-induced modification of the
photon refractive index.

At the interface of two media with different refractive indices 
the transition radiation $\nu\to\nu\gamma$ was studied in \cite{Sakuda:1994zq}
with the assumption of a neutrino magnetic dipole moment.

We presently extend previous studies of the transition radiation
to neutrinos with only standard-model couplings. 
The  media changes the photon dispersion relation. In addition, the
media causes an effective $\nu$-$\gamma$-vertex by
standard-model neutrino couplings to the background electrons. 
We neglect neutrino masses and medium-induced modifications of their  
dispersion relation due to their negligible role.
Therefore, we study the transition radiation entirely
within the particle-physics standard model.

\begin{figure}[b]
  \includegraphics[height=40mm]{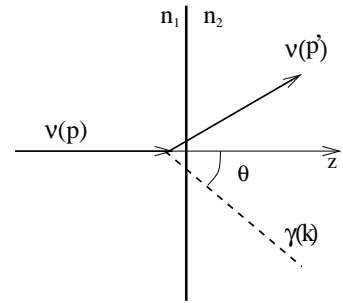}
\caption[...]{Transition radiation by neutrino at an interface of two media with 
refractive indexes $n_1$ and $n_2$.
\label{Fig0}}
\end{figure}

A detailed literature search
reveals that neutrino transition 
radiation has been studied earlier in \cite{Bukina:1998wn}. They used vacuum 
induced $\nu$-$\gamma$ vertex ("neutrino toroid dipole moment") for the $\nu \to \nu \gamma$ matrix element. 
We do not agree with their treatment of the process. 
The media itself induces $\nu$-$\gamma$ vertex. 
 The vacuum induced vertex 
can be treated as a radiation correction to the medium induced one. We found 
that the result of  \cite{Bukina:1998wn}  for the transition radiation rate is more than three orders 
of magnitude, $ ({8\alpha \over \pi})^2$, smaller than our result.  

We proceed by deriving a general expression for the
transition radiation rate (assuming a general
$\nu$-$\gamma$-vertex) in quantum field theory. 
We derive the standard-model
effective vertex in a background of electrons, then we
calculate the transition radiation rate by performing 
semi-analytical integrations and summarize our findings.

{\it Transition Radiation.-}
Let us consider a neutrino crossing the interface of two
media with refraction indices $n_1$ and $n_2$ (see Fig. 1). In
terms of the matrix element ${\cal M}$ the transition radiation
probability  of the process $\nu \to \nu \gamma$ is
 \bea
  W &=&  {1 \over (2\pi)^3}{1 \over 2E \beta_z}
  \int
 {d^3 {\textbf{p}^\prime} \over 2 E^\prime}{d^3
 {\textbf{k}} \over  2 \omega}
 \sum_{pols}\left|\int_{-\infty}^\infty \! \! \! \! \! \! \! dz
 e^{i(p_z-p_z^\prime-k_z)z}{\cal M}\right|^2  \nnb \\
 && \hspace{1cm} \times \ 
 \delta(E-E^\prime-\omega)\delta(p_x^\prime+k_x)\delta(p_y^\prime+k_y) \ .
\label{1}
\eea
Here, $p=(E,{\bf p})$, $p'=(E',{\bf p}')$, and $k=(\omega,{\bf
k})$ are the four momenta of the incoming neutrino, outgoing
neutrino, and photon, respectively and $\beta_z={p_z / E}$. The
sum is over photon polarizations.

We shall neglect the neutrino masses and the deformation of its dispersion relations due to the forward scattering. 
Thus we assume that the neutrino dispersion relation is precisely
light-like so that $p^2=0$ and $E=|{\bf p}|$. 

The formation zone length of the medium is
\be
  |p_z-p_z^\prime-k_z|^{-1}.
\ee
The integral over $z$ in eq. (\ref{1}) oscillates beyond 
the length of the formation zone. Therefore  the contributions to 
the process from the depths over the formation zone length may be neglected. The z momentum $(p_z-p_z^\prime-k_z)$ transfers to the 
media from the neutrino. Since photons propagation in the media suffers from the attenuation(absorption) 
the formation zone length must be limited by the attenuation length of the photons in the media when the later is shorter than the formation zone length.

After integration of (\ref{1})
over $  \textbf p'$ and $z$ we find
 \bea
W &=&  {1 \over (2\pi)^3}{1\over 8E \beta_z}
  \int
 { |{\textbf{k}}|^2 d |{\textbf{k}}| \over \omega
 E^\prime \beta_z^\prime} \ \sin \theta \ d \theta \ d \varphi \nnb \\
 && \times \! \sum_{pols} \left|{{\cal M}^{(1)} \over
 p_z-p_z^{\prime(1)}-k^{(1)}_z} - {{\cal M}^{(2)} \over 
 p_z-p_z^{\prime(2)}-k^{(2)}_z}\right|^2 \! \! , 
 \eea
where $\beta^\prime_z=p'_z / E^\prime$, $\theta$ is the angle
between the emitted photon and incoming neutrino. ${\cal
M}^{(1,2)}$  are matrix elements of the $\nu \to \nu \gamma$ 
in each media. $k^{(i)}_z$ and $p_z^{\prime(i)}$
are $z$ components of momenta of the photon  and of the outgoing 
neutrino in each media. 

 As it will be shown below main contribution to the process comes from large formation zone lenghts and ,thus, small angle $\theta$. Therefore the rate of the process does not depend on the angle betwee the momenta of the incoming neutrino and the boundary surface of two media (if that angle is not close to zero). The integration over $\varphi$ drops out and we may replace $d\varphi \to 2\pi$. $k^{(i)}_z$ and $p_z^{\prime(i)}$ have the forms
 \be
 k_z^{(i)}=n^{(i)}\omega \cos \theta , \ \ 
p_z^{\prime(i)}=\sqrt{(E-\omega)^2-{n^{(i)}}^2\omega^2 \sin^2\theta} \   ,
 \ee
here we have used  $n^{(1,2)}={|\textbf
k|^{(1,2)}}/\omega$.

If the medium is isotropic and homogeneous  the polarization tensor, $\pi^{\mu\nu}$, 
is uniquely characterized by a pair of two polarization functions 
which are often chosen to be the longitudinal and transverse
polarization functions. They can be projected from the full 
polarization matrix. 
In this paper we are interested in transverse photons, since they 
may propagate in the vacuum as well.
The transverse polarization function is 
\be
\pi_t = {1\over2} T_{\mu\nu}\pi^{\mu\nu}, \ 
T^{\mu\nu}=-g^{\mu i}(\delta_{ij}-
{\textbf{k}_i \textbf{k}_j \over {\textbf{k}}^2}) g^{j \nu}.
\ee
The dispersion relation for the photon in the media is the
location of its pole in the effective propagator (which is gauge
independent)
 \be
{1 \over \omega^2 - {\bf k}^2-\pi_t}
 \ee

After  summation over transverse polarizations the 
photons density matrix has a form 
\be
\sum_{trans} \epsilon_\mu \epsilon_\nu =-T_{\mu\nu}= g_{\mu
i}(\delta_{ij}- {\textbf{k}_i \textbf{k}_j \over {\textbf{k}}^2})
g_{j \nu} \ .
 \ee

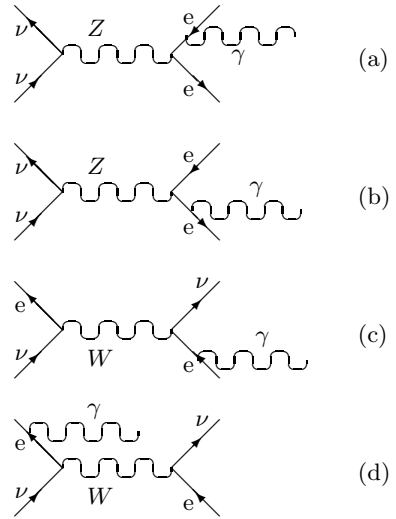
\begin{figure}[!b]
\centering\leavevmode \vbox{ \unitlength=0.8mm
\begin{picture}(60,25)
\put(8,15){\line(-1,1){8}} \put(8,15){\line(-1,-1){8}}
\put(0,7){\vector(1,1){4}} \put(8,15){\vector(-1,1){6}}
\multiput(9.5,15)(6,0){3}{\oval(3,3)[t]}
\multiput(12.5,15)(6,0){3}{\oval(3,3)[b]}
\multiput(33,18)(6,0){3}{\oval(3,3)[t]}
\multiput(30,18)(6,0){3}{\oval(3,3)[b]}
\put(26,15){\line(1,1){8}} \put(34,7){\line(-1,1){8}}
\put(30,19){\vector(-1,-1){1}} \put(28,13){\vector(1,-1){4}}
\put(0,10){\shortstack{{}$\nu$}} \put(12,18){\shortstack{{$Z$}}}
\put(36,14){\shortstack{{$\gamma$}}}
\put(28,8){\shortstack{{e}}}
\put(0,18){\shortstack{{}$\nu$}}
\put(28,20){\shortstack{{e}}}
\put(57,13){\shortstack{{(a)}}}
\end{picture}

\vspace{-0.2cm}

\begin{picture}(60,25)
\put(8,15){\line(-1,1){8}} \put(8,15){\line(-1,-1){8}}
\put(0,7){\vector(1,1){4}} \put(8,15){\vector(-1,1){6}}
\multiput(9.5,15)(6,0){3}{\oval(3,3)[t]}
\multiput(12.5,15)(6,0){3}{\oval(3,3)[b]}
\multiput(31,12)(6,0){3}{\oval(3,3)[t]}
\multiput(34,12)(6,0){3}{\oval(3,3)[b]}
\put(26,15){\line(1,1){8}} \put(34,7){\line(-1,1){8}}
\put(30,19){\vector(-1,-1){1}} \put(28,13){\vector(1,-1){4}}
\put(0,10){\shortstack{{}$\nu$}} \put(12,18){\shortstack{{$Z$}}}
\put(39,15){\shortstack{{$\gamma$}}}
\put(28,20){\shortstack{{e}}}
\put(57,13){\shortstack{{(b)}}}
\put(0,18){\shortstack{{}$\nu$}}
\put(28,8){\shortstack{{e}}}
\end{picture}

\vspace{-0.2cm}

\begin{picture}(60,25)
\put(8,15){\line(-1,1){8}} \put(8,15){\line(-1,-1){8}}
\put(0,7){\vector(1,1){4}} \put(8,15){\vector(-1,1){6}}
\multiput(9.5,15)(6,0){3}{\oval(3,3)[t]}
\multiput(12.5,15)(6,0){3}{\oval(3,3)[b]}
\multiput(32,10)(6,0){3}{\oval(3,3)[t]}
\multiput(35,10)(6,0){3}{\oval(3,3)[b]}
\put(26,15){\line(1,1){8}} \put(34,7){\line(-1,1){8}}
\put(30,19){\vector(1,1){1}} \put(34,7){\vector(-1,1){4}}
\put(0,10){\shortstack{{}$\nu$}} \put(12,9){\shortstack{{$W$}}}
\put(40,13){\shortstack{{$\gamma$}}}
\put(28,8){\shortstack{{e}}}
\put(30,22){\shortstack{{$\nu$}}}
\put(0,18){\shortstack{{e}}}
\put(57,13){\shortstack{{(c)}}}
\end{picture}

\vspace{-0.2cm}

\begin{picture}(60,25)
\put(8,15){\line(-1,1){8}} \put(8,15){\line(-1,-1){8}}
\put(0,7){\vector(1,1){4}} \put(8,15){\vector(-1,1){6}}
\multiput(9.5,15)(6,0){3}{\oval(3,3)[t]}
\multiput(12.5,15)(6,0){3}{\oval(3,3)[b]}
\multiput(4,21)(6,0){3}{\oval(3,3)[t]}
\multiput(7,21)(6,0){3}{\oval(3,3)[b]}
\put(26,15){\line(1,1){8}} \put(34,7){\line(-1,1){8}}
\put(30,19){\vector(1,1){1}} \put(34,7){\vector(-1,1){4}}
\put(0,10){\shortstack{{}$\nu$}} \put(12,9){\shortstack{{$W$}}}
\put(12,24){\shortstack{{$\gamma$}}}
\put(28,8){\shortstack{{e}}}
\put(30,22){\shortstack{{$\nu$}}}
\put(0,18){\shortstack{{e}}}
\put(57,13){\shortstack{{(d)}}}
\end{picture}
}
\caption[...]{Neutrino-photon coupling in  electrons background. \\
 (a,b)~$Z$-$\gamma$-mixing. (c,d)~"Penguin" diagrams
(only for $\nu_e$). 
\label{Fig1}}
\end{figure}

{\it Neutrino-photon vertex-} In a media, photons couple to neutrinos via 
interactions to electrons 
 by the amplitudes shown in Fig 2. One may take into account 
similar graphs with nuclei as well, but their contribution are usually negligible.   
When photon energy is below weak scale ($E \ll M_W$) one may use 
four-fermion interactions and the matrix element for the $\nu$-$\gamma$ vertex  can be written in the form
 \bea \hspace{-0.35cm}
 M \!\! &=&\!\! -{G_F \over \sqrt{2} e} Z \epsilon_\mu
 \bar{u}(p^\prime)\gamma_\nu (1-\gamma_5) u(p) \ (g_V \pi_t^{\mu\nu} \! -
 g_A\pi_5^{\mu \nu})
 \\
  &=&{G_F \over \sqrt{2} e} Z \epsilon_\mu
 \bar{u}(p^\prime)\gamma_\nu (1-\gamma_5) u(p)\ \nnb \\
 && \! \! \! 
 \times
 g^{\mu i}\left(g_V
 \pi_t(\delta_{ij}- {\textbf{k}_i \textbf{k}_j \over {\textbf{k}}^2}) -i g_A \pi_5
 \epsilon_{i j l}{\textbf{k}_l \over |{\textbf{k}}|}\right)g^{j \nu} , 
\eea
here
\bea
 g_V=  \{ \begin{tabular}{l}
        $2 \sin^2 \theta_W +{1\over 2}$ for $\nu_e$
      \\
        $2 \sin^2 \theta_W -{1\over 2}$ for $\nu_\mu, \nu_\tau$
\end{tabular},  \ \ g_A=  \{ \begin{tabular}{l}
        $+{1\over 2}$ for $\nu_e$
      \\
        $-{1\over 2}$ for $\nu_\mu, \nu_\tau$
\end{tabular}.
\eea
$\pi_t^{\mu\nu}$ is the polarization tensor for transverse photons, while  $\pi_5^{\mu\nu}$ is the axialvecor-vector tensor \cite{Braaten:1993jw}.

\begin{figure}[!]
  \includegraphics[height=70mm]{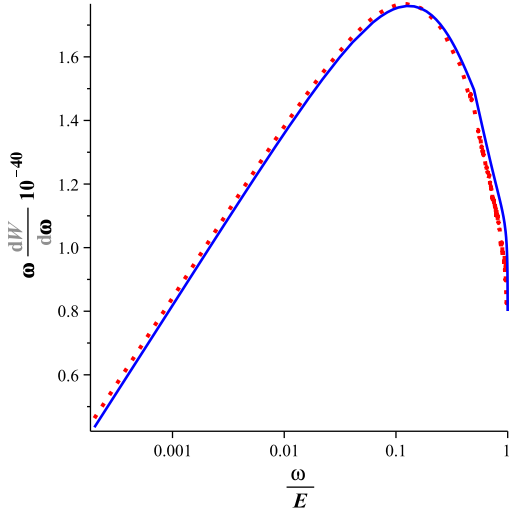}
\caption[...]{Energy spectrum of the transition radiation by electron neutrinos 
at an interface of media with plasma frequency $\omega_p=20$eV and 
vacuum. The energy of the incoming neutrino is $E=1$ MeV. The dot line 
is numerical  and  the solid line is semi-analytical integration over 
angle between photon and incoming neutrino momenta.
\label{Fig3}}
\end{figure}

{\it Transition radiation rate.-}
Armed with these results we may now turn to an evaluation of the 
$\nu \to \nu \gamma$ rate at the interface of the media and the vacuum. 
We find that the transition probability is
 \bea
W &=& {G_F^2   \over 16 \pi^3 \alpha} \int
{\omega \ d \omega \ \sin \theta d\theta  \over (E-p_z^\prime-n
\omega \cos \theta)^2} \ \nnb \\
&& \hspace {-0.5cm} \times
{\Big{[}}( g_V^2
\pi_t^2+g_A^2\pi_5^2)\left (1-{\cos \theta \over
E-\omega}(p_z^\prime \cos \theta - n\omega \sin^2 \theta)\right)
 \nnb
  \\
 && \hspace{-0.0cm}
- 2g_Vg_A\pi_t\pi_5 \left( \cos\theta-{p_z^\prime \cos \theta - 
n\omega \sin^2 \theta \over E-\omega}\right)
{\Big{]}} ,
 \eea
here
\be
p_z^\prime=\sqrt{(E-\omega)^2-n^2\omega^2\sin^2\theta} \ .
 \ee
The maximal allowed angel, $\theta_{max}$ , for the photon emission 
is $\pi/2$ when $\omega<{E \over 2}$ and 
$\sin \theta_{max} ={E-\omega \over \omega}$ when $\omega>{E \over 2}$.

Now we expand underintegral expressions on small angle, since only in 
that case the denominator is small  
(and the formation zone length is large). 
Thus we write the transition probability in the form
 \bea
 W &\simeq&
 { G_F^2   \over 16 \pi^3 \alpha}   \int
 { d \omega \  \theta^2 d\theta^2   
 \over \omega \ 
 [\ \theta^2+(1-n^2)(1-{\omega \over E}) \ ]^2}
 {\Big{[}}( g_V^2
\pi_t^2+g_A^2\pi_5^2)
 \nnb
 \\
 && \hspace{-0.2cm}
 \times(2-2{\omega \over E}+{\omega^2 \over E^2}) -
2g_Vg_A\pi_t\pi_5 {\omega \over E} (2-{\omega \over E}){\Big{]}}
 \ .
 \label{spec1}
 \eea

Eq.(\ref{spec1}) tells us that the radiation is forward peaked 
within an angle of order $\theta \sim \sqrt{1-n^2}$.

After integration over angle $\theta$ we get
 \bea
  W \! &\simeq& \! { G_F^2  \over 16
 \pi^3 \alpha}\! \int \! \! {d \omega \over \omega}  
 {\Big{[}} - \ln [(1-n^2)(1-{\omega \over E})] 
 \label{spectrum} \\
  && \hspace{1.5cm}
  + \ln [(1-n^2) (1-{\omega \over E})  
   +\theta_{max}^2]  \! - 1 \!  {\Big{]}}
 \nnb \\
 && \hspace{-1.2cm} \times 
   {\Big{[}}( g_V^2
 \pi_t^2+g_A^2\pi_5^2)(2-2{\omega \over E}+{\omega^2 \over E^2})-
   2g_Vg_A\pi_t\pi_5 {\omega \over E} (2-{\omega \over E}){\Big{]}}
  . \nnb
 \eea
Numerically eq. (\ref{spectrum}) does not depend much on $\theta_{max}$. 

Usually the axialvector polarization function is much 
less than the vector one. For instance in nonrelativistic and 
nondegenerate plasma these functions are \cite{Braaten:1993jw}
 \be
  \pi_t=\omega_p^2 \  \ {\rm and } \ \   
  \pi_5 ={|{\textbf{k}}|\over 2 m_e} {\omega_p^4 \over \omega^2}
  ,
 \ee
where $\omega_p^2 = {4 \pi \alpha N_e \over m_e}$ is the plasma 
frequency.
Therefore we may ignore the term proportional to $\pi_5$.

Since we are interested in the forward radiation in the gamma ray region, we assume that the index of refraction of the photon is
 \be
 n^2=1-{\omega_p^2 \over \omega^2} 
  \ee
and the photons from the medium to the vacuum propagate without any reflection or/and refraction.

In Fig.3 we plot the energy spectrum of the photons from the transition radiation by electron neutrinos with energy $E=1$MeV. 

After integration over photon energy we find  the neutrino transition radiation 
probability as
 \be
W = \int^E_{\omega_{min}} \! \! \! \! \! \! \! \! \! dW \simeq {g_V^2 G_F^2 \omega_p^4 \over 16
\pi^3 \alpha} \left ( 2 \ \ln^2 { E \over \omega_p}-5 \ln {E \over \omega_p}+\delta \right) 
 \label{rate} 
\ee
here $\delta \simeq 5$ for $\omega_{min}=\omega_p$, $\delta \simeq -1$ for $\omega_{min}=10 \omega_p$.

The energy deposition of the neutrino in the media due to the transition radiation 
\be
\int^E_{\omega_p}\! \! \! \!  \omega \ d  W_{\nu \to \nu \gamma}  \simeq {g_V^2 G_F^2 \omega_p^4 \over 16
\pi^3 \alpha} E \Big{[}{8 \over 3} \ln {E \over \omega_p}- 4.9 + 9 {\omega_p \over E} + O({\omega_p^2 \over E^2}) \Big{]}
\label{ed}
 \ee

The eqs.(\ref{spectrum}),(\ref{rate}) and (\ref{ed}) are main results of the present work.

For MeV electronic neutrinos the transition radiation probability is about  $W\sim10^{-39}$ and the energy deposit is about 
$1.4 \cdot 10^{-34}$ eV  when they cross the interface of the media with  $\omega_p=20$ eV to vacuum. 

Unfortunately the transition radiation probability is extremely small and cannot be observed at the Earth. 

On the other hand at the surface of the neutron stars electron layer may exist with density $\sim  m_e^3$, due to the fact that the electrons not being bound by the strang interactions are displaced to the outside  of the neutron star \cite{PicancoNegreiros:2010uc}.  Therefore MeV energy neutrinos emited by the neutron stars during its cooling processes will have transition radiation with the probability of $\sim 10^{-20}$ and energy spectrum given in eq. (\ref{spectrum}).

{\it Summary and Conclusion-}
We have calculated the neutrino transition radiation at the interface 
of two media. The charged particles of the media provide an effective 
$\nu$-$\gamma$ vertex, and they modify the photon dispersion relation. 
We got analytical expressions for the energy spectrum of the transition radiation,  its probability and the energy deposition at the process.
The radiation is forward peaked within an angle of order 
${\omega_p \over \omega}$. The photons energy spectrum is falling almost linearly 
 over photon energy. Close to its maximum value ($\sim E_\nu$) it is further suppressed 
due to the smaller phase space.

{\it Acknowledgments-} We are grateful to K. Ispirian for bringing our attention to \cite{Bukina:1998wn} and for discussions. 
The research was supported by the Schwinger Foundation and by the RGS Armenia.


\end{document}